# Fog Assisted Cloud Models for Smart Grid Architectures- Comparison Study and Optimal Deployment

Md. Muzakkir Hussain[1] , Mohammad Saad Alam[2] , M.M. Sufyan Beg[1]
1Department of Computer Engineering, ZHCET, AMU
md.muzakkir@zhcet.ac.in
mmsbeg@cs.berkeley.edu
2Department of Electrical Engineering, ZHCET, AMU
saad.alam@zhcet.ac.in

***Abstract*— Cloud Computing (CC) serves to be a key driver for fulfilling the store and compute requirements of a modern Smart Grid (SG). However, since the datacenters are deployed in concentrated and far remote areas, it fails to guarantee the quality of experience (QoE) attributes for the SG services, viz. latency, bandwidth, energy consumption, and network cost. Fog Computing (FC) extends the processing capabilities into the edge of the network, offering location-awareness, low latency, and latency-sensitive analytics for mission critical requirements of SG. In this work, we first examine the current state of cloud based SG architectures and highlight the motivation(s) for adopting FC as technology enabler for sustainable and real-time SG analytics. Then we present a hierarchical FC architecture for supporting integration of massive number of IoT devices into future SG. Following this architecture we proposed a cost optimization framework that jointly investigates data consumer association, workload distribution, virtual machine placement and QoS constraints towards viable deployment of FC model over SG networks. The formulated MINLP problem is then solved using Modified Differential Evolution (MDE) algorithm. Comprehensive evaluation of proposed framework on real world parameters shows that for an infrastructure with nearly 50% applications requesting real-time services, the overall service latency for fog computing get reduced to almost half of that of generic cloud paradigm. It is also observed that the fog assisted cloud framework lowers the aggregated electricity consumption of the pure cloud computing paradigm by more than 40%.**

## NOMENCLATURE

| SYMBOLS | DEFINITION |
|---|---|
| T | Length of time frame |
| $\pi_C$ | Probability that a packet is directly dispatched to cloud |
| $\pi_F$ | Probability that a packet is pushed to fog |
| $\omega_{jk}$ | Bandwidth of Consumer-Fog link |
| $\boldsymbol{\delta}$ | Data rate of bandwidth unit/sub carrier allotted |
| $\alpha_{comm}$ | Delay to cost conversion parameter |
| $r_i^a$ | Traffic arrival rate |
| $r_i^s$ | Service Rate |
| $\ell_c$ | Length of Data demanding multiple fog hops |
| $r_{f-f}$ | Data rate of Inter-fog communication links |
| $\chi_{f-c}$ | Delay of WAN transmission path |
| $\Gamma_{f-c}$ | The traffic rate dispatched from the fog device (no. of requests |
| | per second) *i* to the cloud server *j* |
| $n_{m,i}$ | No. of turned-on machines in server *i* |
| $\Gamma_i$ | Traffic arrival rate of machine *i* at cloud server |
| $\eta$ | CPU frequency at cloud server i |
| $K$ | No. of cycles per request |
| $\Theta_{i,j}$ | Denotes the traffic between clusters (VM) *i* and *j* |
| $\mho_{f,f'}$ | Cost of unit traffic between clusters (VM) *i* and *j* |
| $\alpha_{cons}$ | Power to cost conversion parameter |
| $p_{a-f}$ | Energy required to upload unit byte of data-stream |
| $p_{f-f'}$ | Energy required to transmit unit byte fog-fog of data-stream |
| $p_{f-c}$ | Energy required to dispatch unit byte of data-stream (fog -cloud) |
| $p_f^{comp}$ | Average power consumption per byte of Computation |
| $\psi_f$ | Weight-factor associated with the data-set required for analysis |
| $a_i$ $b_i$ $c_i$ $A_i$ $B_i$ | Positive architectural constants |
| $\zeta$ | Cost of carbon footprint |
| $R$ | average carbon emission rate |
| $PUE$ | Power usage effectiveness |
| $\beta_c$ | Server power at the cloud |
| $\lambda_{f-f'}^a$ | Inter-fog arrival data rate |
| $\mathbb{R}$ | Arbitrary large real number |
| $\mathcal{E}^a$ | Scaling factor |
| $Q_f$ | Capacity of storage resources of fog node *f* |
| $BV_1$ | Binary variable deciding cloud/fog scale offloading |
| $BV_2$ | Binary variable deciding if fog *f* is associated to consumer *d* |
| $BV_3$ | Binary variable indicating if a bandwidth units allocated to consumer |
| $BV_4$ | A binary variable indicating if data from fog node *f* for application *a* is associated in fog node $f'$ |
| $BV_5$ | A binary variable indicating if a VM for application a is placed in fog node $f'$ |
| $BV_{c,i}$ | Denotes the on/off state of cloud server *i* |
| $q_t^I$ | State of Input Buffer |
| $q_t^{I,d}$ | Size of Workload Drained from Input Queue |
| $q_t^O$ | State of Output Buffer |
| $q_t^{O,d}$ | Size of Workload Drained from output Queue |
| $VM_M^I$ | Maximum number of VMs hosted by FCN *i* |

## ACRONYMS

| | |
|---|---|
| SG | Smart Grid |
| CC/FC | Cloud/Fog Computing |
| VM | Virtual machines |
| BU | Bandwidth Units |
| EV | Electric Vehicles |
| IoT/IoE | Internet of Things/Energy |
| P2P | Peer to Peer |
| M2M | Machine to Machine |
| ISO | Independent System Operators |
| RSU | Road-Side Units |
| RTU | Remote Terminal Unit |
| QoS/QoE | Quality of Service/Experience |
| MDE | Modified Differential Evolution |
| CRB | Computational Resource Blocks |
| AMI | Advanced Metering Infrastructure |
| PMU | Phasor Measurement Unit |
| CAP | Consistency Availability Partition |
| SOA | Service Oriented Architecture |
| MDC | Mega Data Centers |
| WAN | Wide Area Networks |
| ACR | Admission Control Router |
| FCN | Fog Computing Node |
| FNE | Fog Network Efficiency |
| BDA | Big Data Analytics |
| RFID | Radio-Frequency Identification |
| HVAC | Heating, Ventilation and Air-Conditioning Systems |
| I/P/S/FaaS | Infrastructure / Platforms /Software/ Fog as a Service |
| VANET | Vehicular Ad hoc Network |

# I. INTRODUCTION

A Smart Grid is a pervasive network of densely distributed energy and resource-limited wireless things (e.g., smart devices), all capable of gathering and transferring in real-time large volumes of heterogeneous environmental data. However, due to the current energy-computing-bandwidth limitations of the wireless domain, up to date, a system of this complexity was unfeasible [1]. Through the notion of IoT, the upcoming Future Internet is bringing the SG machines, devices and sensors connected to the internet [2], [3]. By interconnecting the machines with the internet they become smart, with the ability to react and make decisions on their own. IoT devices are connecting wirelessly or directly through network switches and devices. Many of the devices have closed interfaces that make it hard to extract information. Some devices are old, some are new, and all of them communicate with different protocols, which mean a lot of customization from different vendors and expensive professional services firms. With machines and sensors recording data in real time every second or minute, the amount of data can be massive. Latency can be a big issue when real-time decisions need to be made [2], [4]–[7]. When multiplied by hundreds of machines across multiple plants the amount of data can be too much for the network to handle when it finally reaches the cloud data center. Finally trying so many machines and devices together means exorbitant customization costs. In such circumstances, reaching a consensus on where to install the compute and storage resources becomes a critical research thrust for the academia, industries, R&D and legislative bodies [8]. The centralized CC model, though have virtually unlimited resource pooling capabilities, it ceases to welcome its proponents because of its failure in building common and multipurpose platform that can provide feasible solutions to the mission critical requirements of an IoT aided SG [9].

FC is a distributed computing framework installed on the intermediary network switches, devices sensors and machines to eliminate the connectivity, bandwidth and latency issues prevalent in CC. FC is a natural extension of CC and is foreseen as a remedy to alleviate such issues [10]. Both these platforms complement each other to form a mutually beneficial and inter-dependent service continuum between the Cloud and the endpoints to make computing, storage, control, and communication possible anywhere along the continuum. A robust FC based SG topology will allow dynamic augmentation of associated fog nodes (FCNs), thus rejuvenating the elasticity and scalability profiles of mission critical SG requirements. The objective of FC is to standardize and ensure secure communication between all devices, SG components, old and new, across all vendors, so that customization and service costs are kept to a minimum. The FCNs sit on or between devices and the cloud. The notion is to allow machines to speak directly to each other without going through cloud by connecting machines, devices and sensors directly to each other (at only one or few steps) enabling realtime decisions to be made without transmitting vast amount of data through the cloud. The FC objective is also to connect all devices to the cloud with open communication standards [11]. The result is a smart network of devices that are able to make decisions themselves and react in realtime to a changing environment or supply chain such as SG. Further it will: (i) Make electric utilities smarter. (ii) Help accelerate the transformation of cities into smart cities. (iii) Accelerate industries into smart manufacturing. (iv) Make SG logistics even more efficient and etc.

Since the FC elements are heterogeneous and resource constrained, proper orchestration among them is very essential to attain a near optimal performance. Moreover, the workload allocation in FC should be performed so that the resource utilization rate is maximized. In other words, for efficient execution of FC, the resources should be provisioned to guarantee an optimal balance of attributes defining architectural QoS (e.g. power consumption, carbon footprint etc) and QoE (near real-time response) of user. As the FCNs exhibit diverse cost profiles, it is necessary to have an optimal user-to-FCN association such that the corresponding upload latency is minimized. Also, the number of successful connections is limited by the number of Bandwidth Units (BU) or computational resource blocks (CRBs) available at each FCN. For each workload to be allocated, finding a proper set of FCNs to host the VMs for each application is also a key issue to cost minimization. Therefore, in this paper, we are motivated to investigate the QoS guaranteed minimum cost resource management problem in fog enabled SG architectures.

The main goals and organization of this paper is as follows. In section II, we synoptically revisit the current state of cloud based SG. Correspondingly, we also examine how the FC can serve as an ally to cloud computing platforms and how far such a noble mix of both computation models will successfully satisfy the high assurance and mission critical computing needs of SG. In section III, a hierarchical FC architecture is introduced, describing its working in the context of SG ecosystems. Section IV presents a mathematical framework for defining the cost profiles in both cloud and fog computing. Consecutively, a cost-efficient optimization framework for cumulative assessment of user to FCN association, workload allocation, and VM placement constraints, is proposed towards viable deployment of FC. Further, a Modified Differential Evolution (MDE) is employed in section V to solve this model. Section VI presents a comprehensive cost comparison of FC over generic cloud computing technique through attributes contributing metrics, viz. Latency cost, Energy Consumption cost, etc. In section VII, the related works unveiling the viability of FC based SG architectures are presented along with a brief discussion of the state-of-the-art work in the area of the resource provisioning and workload allocation in FC. Section VIII concludes the manuscript by providing insights, challenges and future prospects towards sustainable FC based SG deployments.

## II. Smart Grid As Use-Case for Fog Computing

The FOG computing (FC) paradigm will potentially abridge the silos between personalized and batch analytics in SG informatics. The FC is an architectural setup for federated as well as distributed processing where application specific logic is embedded not only in remote clouds or edge systems, but also across the intermediary infrastructure components. A robust fog topology allows dynamic augmentation of associated fog nodes, thus significantly improvising the elasticity and scalability profiles of mission critical infrastructures. In this section, we outline the fog computing paradigm and examine its preeminence over the cloud computing in SG context. We highlight some key SG characteristics that motivate the analytics utilities to relinquish the current cloud adoption and opt for analytics at the edges of the SG networks.

*A. Decentralization and Low Latency Analytics:*

The data generation and consumption sites of SG entities and stakeholders are sparsely distributed all over the SG network. Centralized cloud model succumbs to serve as the optimal strategy for SG applications that are geographically distributed. For instance, together with a centralized control, the sensor and actuator nodes deployed across the smart home or EV networks also require geo-distributed intelligence. The domain of information transparency may need to be extended from mere SCADA systems to a scale that ensures national level visibility. However, since majority of the SG services are consumer centric, they demand location aware analytics to be performed closer to the source of the data.

The contemporary cloud infrastructures pose serious latency issues for power applications operated by real-time decisions. For instance, the SCADA system employed in a modern data driven SG is so timed that it may produce glitches when operated over ubiquitous TCP/IP protocols. Fortunately, the FC is there to harness the store and compute resources latent in the underutilized SG resources such as vehicles [12], [13], gateways, PMUs etc [11]. Fog model complements the cloud computations with dedicated and ad-hoc computational resources [14], to be performed on the edge nodes of an IoT aided SG thereby reducing the networking latency.

*B. Limited resource distribution for individual IoT endpoints:*

Compared to mega servers in cloud computing, each individual IoT devices such as sensors and actuators have limited store and process capabilities. The front end mobile devices may fail to perform complex SG analytics due to hardware restrictions such as draining of battery, or other middleware limitations. Often the data needs to be sent to cloud to meet the processing demands and meaningful information is then relayed back to the front-end [15]. But if carefully designed, such resource scraps may be aggregated and utilized for dedicated purposes. Think of a smart vehicle having limited processing and storage resources, thanks to the parked vehicular cases where these underutilized resources can be aggregated, utilized to perform alluring services such as internet of vehicles (IoV), Social internet of vehicles (SIoV), infotainment services etc. Moreover, for many of the SG usecases, not all data from a front-end device will need to be used by the service to construct analytical workloads on the cloud. Potentially, data can be filtered or even analyzed at such fog nodes equipped with spare computational resources, to accommodate data management and analytics tasks.

*C. Energy Consumption of Cloud Data Centers:*

The energy consumption in mega data centers is likely to get tripled in in coming decade. Adopting energy aware strategies becomes an earnest need for computational folks. Offloading the whole of smart grid applications into the cloud data centers causes untenable energy demands, a challenge that can be alleviated by adopting sensible energy management strategies. There are plenty of SG applications that can be run without significant energy implications. For such trivial services, instead of overloading data centres, the analytics can be made ready at SG fog nodes such as RTUs, SCADA systems, roadside units (RSU), base stations and network gateways.

*D. Handling Data Deluge and Network Traffic:*

The population IoT endpoints in SG architectures is growing at an enormous rate, as can be discovered from the smart meter installation landscape in [16]. As an illustration, consider a smart meter that is reporting data at frequency 4 times per hour. It will generate 400MB of data a year. Thus, a utility serving AMI to a million customers will generate 400TB of data a year. In 2012, BNEF predicted 680 million smart meter installations globally by 2017—leading to 280PB of data a year. This is not the only data utilities are dealing with, the generated data volume is anticipated to come from other SG attributes such as consumers load demand, energy consumption, network components status, power-line faults, advanced metering records, outage management records and forecast conditions etc. Electric Power Research Institute (*smartgrid.epri.com*) estimates the exponential boom in the quantities of smart grid data for a vertically integrated utility serving about one million customers [16].

One solution to cope with such Big Data avalanche is to have an expansion of data center networks that can mitigate the analytical workloads. However, this again raises concerns related to sustainable energy consumption and carbon footprint. Attempts to undertake analytics on the edge device is restrictive due to their resource limitations. Also in many cases, aggregated and collective analytics becomes unfeasible. Added to this is the volume of network traffic and complexity in SG that worsens the reliability and availability of analytics services. Leveraging the SG architectures with dedicated fog nodes deployed at a few hop distances (mostly one hop) from the core network, will complement the computations of both front end device as well as back-end data center.

### *E. Security Concerns:*

In the context of SG applications security is defined to mean the safety and stability of the power grid, rather than protection against malice, as the latter comes under the privacy umbrella. The malignancy of casual justification of CAP theorem is manifested in the current position of SG cloud security [17]. In current SG designs, the whole data find its way into cloud storages that comprises of huge number servers and storage elements having peculiar horizontal and vertical elasticity. Unfortunately, the existing cloud based security and privacy enforcements are precisely erratic, and many a times the cloud operators may turn to be evil if they are intended to do so. In a competitive plus shared cloud SG environments the worry is that the rivals may spy on property data, leading to cyber-physical war (CPW) in worst cases.

Gartner claimed that the cloud platforms are fraught with security risks and suggests SG like customer must put rigorous questions and specifications before the cloud service providers [18], [19]. They should also consider a guaranteed security assessment from a neutral third party prior to making any commitment. The woeful protection services of current cloud deployments often stimulate the cloud vendors to recapitulate their security management folks to “not be evil”. Rigorous efforts are on headway across the power system and transportation communities to come up with SG cloud utilities and platforms leveraged with robust protective contrivances where the stakeholders could entrust the storage of sensitive and critical data even under concurrent share and access architectures [20].

The distinguishing geo-distributed intelligence provided by FC deployments make it more viable for security constrained services as the critical and sensitive tasks are selectively processed on local fog nodes and are kept within the user control, instead of offloading the whole universe of datasets into the vendor regulated mega data centers.

## III. FOG COMPUTING ARCHITECTURE FOR SMART GRID

A SG offers a rich use-case of fog computing. Consider an IoT aided SG architecture, where we have a large-scale, geographically distributed micro-grid (e.g. wind farm) system populated with thousands to millions of sensors and actuators. This system may further consist of a large number of semi-autonomous modules or sub-systems (turbines). Each subsystem is a fairly complex system on its own, with a number of control loops. Established organizing principles of large-scale systems (safety, among others) recommend that each subsystem should be able to operate semi-autonomously, yet in a coordinated manner. For that, controller with global scope, implemented in a distributed way may be employed. The controller builds an overall picture from the information fed from the subsystems, determines a policy, and pushes the policy for each subsystem. The policy is global, but individualized for each subsystem depending on its individual state (location, wind incidence, conditions of the turbine). The continuous supervisory role of the global controller (gathering data, building the global state, determining the policy) creates low latency requirements, achievable locally in the edge centered deployment also known as the Fog. Such system generates huge amounts of data, much of which are actionable in real time. It feeds the control loops of the subsystems, and is also used to renegotiate the bidding terms with the ISO whenever necessary. Beyond such real-time network applications, the data can be used to run analytics over longer periods (months, years) and over wider scenarios (including other wind farms or other energy data). The cloud is the natural place to run such batch analytics. The SG requires a store and computing framework leveraged with efficient communication network connecting the subsystems, the system and the Internet at large (cloud).

The underlying notion of fog computing is the distribution of store, communicate, control and compute resources from the edge to the remote cloud continuum [3]. The fog architectures may be either fully distributed, mostly centralized, or somewhere in between. In addition to the virtualization facilities, specialized hardware and software modules can be employed for implementing fog applications. In the context of an IoT aided SG, a customized fog platform will permit specific applications to run anywhere, reducing the need for specialized applications dedicated just for the cloud, just for the endpoints, or just for the edge devices. It will enable applications from multiple vendors to run on the same physical machine without reciprocated interference [9]. Further, the fog computing will provide a common lifecycle management framework for all applications, offering capabilities for composing, configuring, dispatching, activating and deactivating, adding and removing, and updating applications [21]. It will further provide a secure execution environment for fog services and applications [3].

A multi-tier fog assisted cloud computing architecture is shown Fig 1, where a substantial proportion of smart grid control and computational tasks are non-trivially hybridized to geo-distributed fog computing nodes (FCN) alongside the data center based computing support. The hybridization objective is to overcome the disruption caused by the penetration of IoT utilities into SG infrastructures that calls for active proliferation of control, storage, networking and computational resources across the heterogeneous edges or end-points. The framework facilitates the comprehensive enactment of IoT services in a fog landscape supporting big data analytics of SG data and guaranteeing optimal resource provisioning in the fog. The lowermost tier consists of smart grid infrastructure populated IoT devices, that are further meshed by non-invasive, highly reliable, and low cost sensory nodes, deployed across the SG horizontals i.e. generation, transmission, distribution and consumption [21]. Thus, the physical components of this tier comprises of RFIDs, devices, cameras, infrared sensors, laser scanners, GPSs and miscellaneous data collection entities.

The next tier also called fog computing tier functions as the prime component of a typical FC model. Though the notion of fog computing node (FCN) were mainly proposed by Cisco in [22] and Bonomi [23], the distribution, computational and storage capacities, their interaction and their deployment as a service (FaaS) scheme has not been clearly classified. The FCNs are expected to analyze the datasets based on the non-functional requirements of supporting applications such as latency, QoS, reliability etc. Further, the massive sensing data streams generated from these geospatially distributed sensors have to be processed as a coherent whole.

The fog computing layer in Fig. 1 (within dashed oval) integrates the intermediate computing services into various sub-layers. The lowest layer, closest to the physical layer comprises of multiple low-power and high-performance computing nodes or edge devices such as dedicated routers and cellular network base stations etc. Each edge device covers a group of sensors in its domain for performing traces of local and instant analytics. The outputs of edge devices may either fully assimilated within the SG applications or may be offloaded to the upper tier for further processing. The later may be reports of accomplished tasks or some pre-processed datasets that are made ready for upper level analysis. For example, the instant output can be used to provide real-time feedback control to a local infrastructure e.g. to inform police authorities in response to any isolated and small threats to a monitored electric vehicular network.

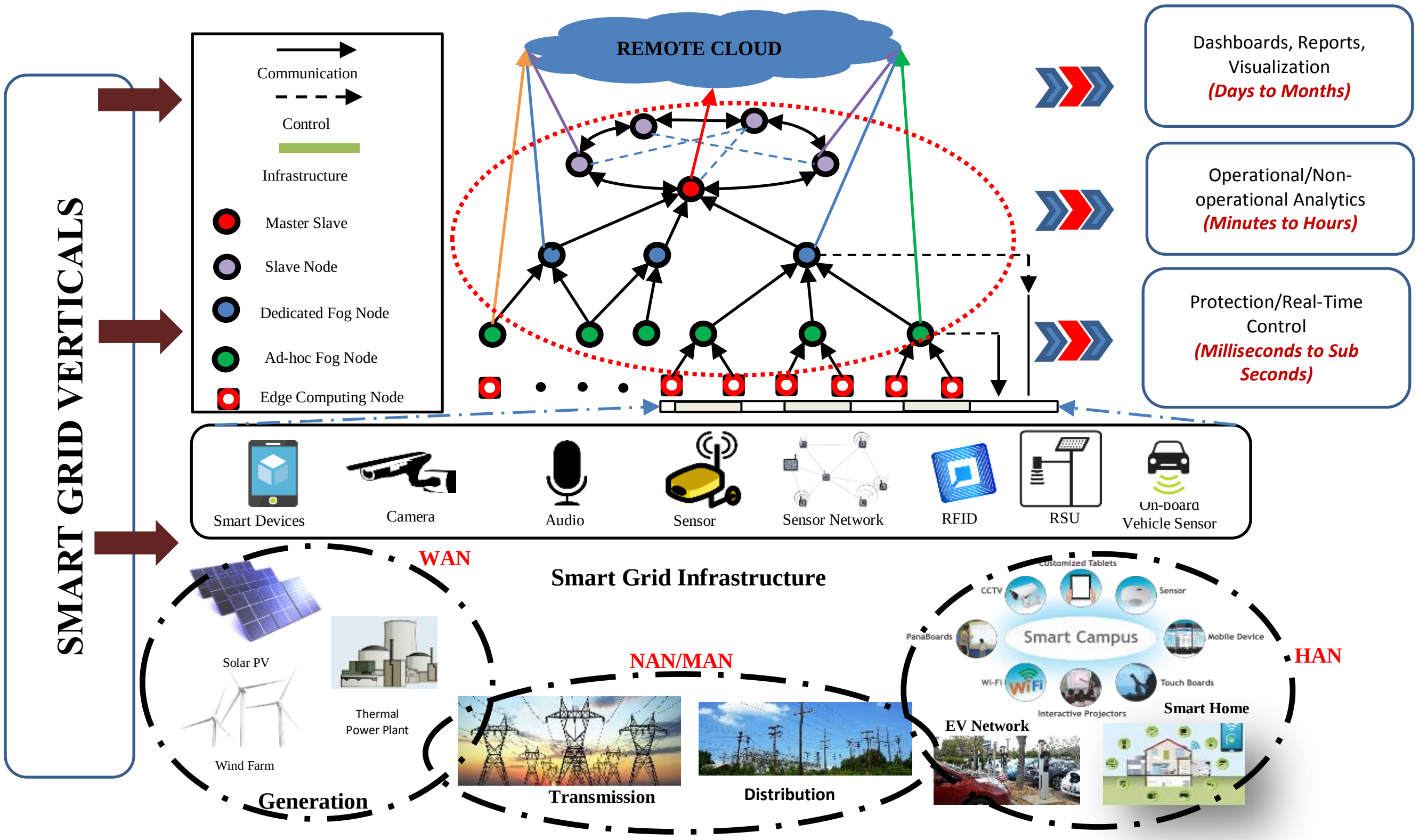

**Figure 1: Topology of Fog Architecture in IoT aided Smart Grid**

The higher sub-layer consists of dedicated computing nodes named fog computing nodes (FCN), either connected to edge nodes from lower layers or to upper layer cloud data centers through reliable communication links. Sometimes, the FCNs at same depth are paralleled to nodes lying below in the hierarchy to undertake tasks. In many cases the FCN may form further sub-trees of FCNs, with each node at higher depth in the tree managed by the ones at lower depth, in master slave paradigm [32]. A typical association of such hierarchies is depicted fog computing layer of Fig.1. To be specific, consider a SG power transmission scenario where the FCNs are assigned with spatial and temporal data to identify potential hazardous events in transmission lines viz. power thefts, intrusion in the network etc. In such circumstances, these computing nodes will shut down the power supply from the distribution sub-station and the data analysis results will be feed backed and reported to the upper layer (from village substation to SCADA, to city wise power distribution center or to generation bodies) for complex, historical and large-scaled behavioral analysis and condition monitoring. The distributed analytics from multi-tier fogs (followed by aggregation analytics in many use-cases) performed at the fog computing layer acts as localized "reflex" decisions to avoid potential contingencies. Meanwhile, a significant fraction of generated IoT data from smart grid applications don't require to be dispatched to the remote clouds, hence response latency and bandwidth consumption problems could be easily solved.

The uppermost tier is the Cloud Computing or data center layer, providing global or centralized monitoring and control. The data centers are leveraged with high performance distributed computing and storage elements that allow to perform complex, long-term (days to years), and grid-wide behavioral analysis. The results of cloud scale analytics may be large-scale event detection, long-term pattern recognition, and relationship modeling, to support dynamic decision making. One major objective of cloud level analytics is to ensure the grid and service vendors to perform large scale resource and response management activities and to be prepared for blackouts or brownouts.

## IV. NETWORKING AND SYSTEM FORMULATION

In a cloud computing model the mega data center (MDC) provides sharable resource pool available for on-demand use. Since the MDCs are far away from the data generation sites, data migration and service latencies give rise to infeasibilities for real-time and interactive SG applications. However, in fog architecture, low/battery powered FCNs are deployed at the dedicated edges of the network to offer store, compute and networking support for SG mobility, real-time response and geo-distributed intelligence. Consider an FC architecture customized to BDA in SG applications, in which data and computation are selectively offloaded to either cloud or fog scale processing, guided by an application specific logic. Without loss of generality, let us assume the sets D , F , and N represents the set of data centers, fog nodes and number of consumers having cardinality ***D***, ***F*** and ***N*** respectively. An instance of Smart Grid (SG) network can be modeled as a connected cellular graph of order (***N*** +***F***) whose vertices are constituted by data consumer set (N) and FCN set (F). Let $r_i^a$ be the frequency of workload arrivals on fog node *i*. For simplicity it is assumed that the FCNs are equipped with homogenous processing elements (E) each having service rate $r_i^s$. An FCN *j* is reachable from query source node *k* if *j* is in the preference list *L*, dynamically maintained by FCN *j*.

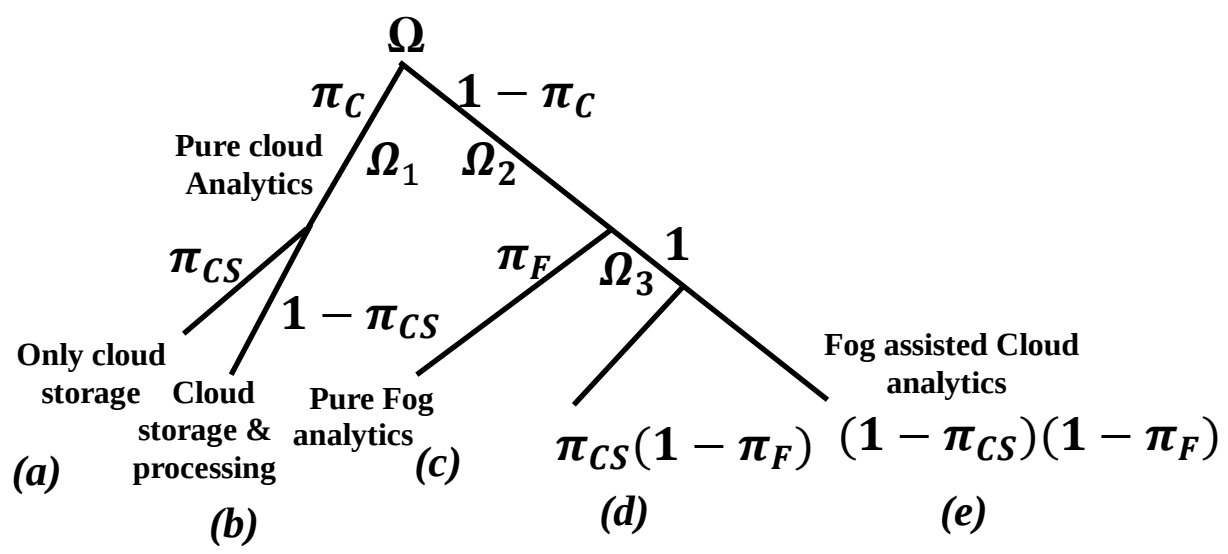

**Figure 2: Probabilistic Decision tree depicting the Workload Offloading Strategy**

Consider a pilot SG analytics service to be delivered from the hierarchical fog architecture shown in Fig. 1. Over the time frame considered, out of volume Ω of generated workload, the sensing and offloading schemes directly dispatch (towards the left sub-tree of root in Fig. 2) the less critical datasets $\Omega_1$ (demanding historical analytics on power market operational data, forecasting data etc.) for cloud scale processing. The latency critical datasets $\Omega_2$ are uploaded to the associated FCNs with probability $1-\pi_c$. A fraction of datasets ($\Omega_3$) demand sequential execution of both cloud and fog scale algorithms, where the results of fog analytics are used for reflex and real-time decision making, and consecutively dispatched to remote clouds for operations such as large scale event detection, behavioral analysis, prediction and pattern analysis etc. The uncertainties associated with such multi-modal execution are captured through probability terms appearing in the leaves ((a) to (e)) of decision tree in Fig. 2. An ideal fog-cloud framework is leveraged with robust inferencing logic and intelligent filtering devices to undertake instant decisions on where to distribute the produced datasets. Following these assumptions, in this section, we establish a mathematical framework for defining the cost profiles in both cloud and fog scale processing. The objective of the proposed framework is to minimize the overall cost incurred due to power consumption, latency and carbon footprint.

*A. Cost profile for Generic Cloud processing:*

Equation (1) represents the total cost of a traditional data center based computational infrastructure. The addend terms are respectively source to cloud communication latency cost, cost of storage and analytics, cost corresponding to electricity consumption and cost reflecting the volume of carbon footprint from these data centers.

$$C_C^T = C_C^{comm} + C_C^{comp.} + C_C^{cons.} + C_C^{ems.} \qquad (1)$$

*B. Cost profile for Fog assisted Cloud processing*

Equation (2) denotes the expected cost of a fog assisted cloud architecture. The first term gives the cost corresponding to the edge level processing for data demanding real-time or time critical processing. An optimal workload allocation algorithm filters the datasets which are less latency sensitive or that demand bulky resources and offloads them directly to the cloud level processing logic. Thus the cost profile of second term in (2) is similar to generic cloud analytics as shown in (1). The uncertainty associated with workload offloading is captured through variable $\pi_c$ shown in Figure 2.

$$C_F^T = \sum_{\substack{t=0 \\ \forall i \in I^+}}^{T} C_i^t = \sum_{t=0}^{T} (1-\Pi_C).BV_1 C_{f,i}^t + \Pi_C.(1-BV_1) C_{c,i}^t \qquad (2)$$

Without loss of generality, the cost term corresponding to fog level processing in (2) is given by:

$$C_{f,i}^t = C_{f,i}^{comm} + C_{f,i}^{comp.} + C_{f,i}^{cons.} + C_f^{ems.} \qquad (3)$$

where, the addend terms are analogous to (1), i.e. cost due to fog communication latency, cost of computation, cost of power consumption and the cost corresponding to carbon footprint in fog assisted cloud network. The cost of communication $C_{f,i}^{comm}$ in (4) includes the overheads corresponding to data upload delays from user applications to FCNs, processing latency at candidate FCNs, and inter-fog communication delays. It also includes latency cost in fog to cloud dispatch and price of cloud level batch analytics.

$$C_{f,i}^{comm} = \left\{ \begin{array}{l} L_{f,i}^{upd} + L_{f,i}^{comp.} + \Pi_F.L_{f-f,j}^{comm.} + (1-\Pi_F).L_{f-c}^{dispatch} + \\ (1-\Pi_F).(1-\Pi_{CS}) L_c^{comp} \end{array} \right\}.\alpha_{comm.} \qquad (4)$$

Equations (5)-(9) respectively define the each cost term involved in (4). For application(s) associated to fog node *k* via link *j* of bandwidth *w*, the upload latency is given by (5). Assuming a queuing system, for the fog device *i* with the traffic arrival rate $r_i^a$ and service rate $r_i^s$, the computation latency is given by (6) that involves queuing delay while waiting for service, service time and time taken for VM installation. The equation may also be extended to capture the scenario when the task (depending upon the size of data) needs to be paralleled to multiple fog nodes. We omit additional latency term due to aggregation of results from any such concurrent computations. Often, the data passes through multistage fog computations involving inter-fog communication delay, given by (7). A significant fraction of data coming out from the fog network also needs to be dispatched to the cloud for permanent storage and historical analysis. The delay due to transmission of data over the fog to cloud WAN transmission link is captured in (8).

$$L_{f,i}^{upd} = \frac{\Omega_2}{\omega_{jk}} = \frac{\Omega_2}{\sum_{ch \in CH} BV_3.\delta} \quad (5)$$

$$L_{f,i}^{comp.} = 1 \Big/ \left( r_{i,f}^{s} - r_{i,f}^{a} \right) \quad (6)$$

$$L_{f-f,i}^{comm.} = \frac{\ell_c}{r_{f-f}} \quad (7)$$

$$L_{f-c}^{dispatch} = \chi_{f-c}.\Gamma_{f-c} \quad (8)$$

$$L_c^{comp} = W\left( n, \frac{r_{i,c}^{s}}{r_{i,c}^{a}} \Big/ (r_{i,c}^{a} - r_{i,c}^{s}) \right) + \frac{1}{r_{i,c}^{s}} \quad (9)$$

The processing in cloud data server is characterized through an ***M/M/n*** queuing model having cloud response time of the order given by (9), where $W(n/\lambda)$ is the Erlang's C formula [24], $\lambda$ is the computational performance index defined as the ratio of traffic arrival rate to the service rate. If VM *i* and *j* are installed into fog nodes *f* and *f'* respectively, the aggregate traffic cost or the cost of computation $C_f^{comm}$ is given by:

$$C_f^{comp.} = \sum_{c_f} \sum_{c_{f'}} BV_f^{i}.BV_{f'}^{j}.\Theta_{i,j}.\mho_{f,f'} \quad (10)$$

The net power consumption term in a typical FC model involves energy expended due to transmission of byte stream from data generation nodes to cloud data center(s) via fog node(s) and due to computations across FCNs and data centers. Thus $P_{net,t}^{cons}$ is given by:

$$\begin{aligned} P_{net,t}^{cons.} &= P_{aff\dot{c},t}^{comm.} + P_{t}^{comp.} = P_{aff\dot{c}}^{comm.} + P_{f}^{comp.} + P_{c}^{comp.} \\ &= \left[ P_{a-f}^{cons.} + P_{f-f'}^{cons.} + P_{f-c}^{cons.} \right] + P_{f}^{comp.} + P_{c}^{comp.} \\ &= \begin{bmatrix} BV_2.p_{a-f}.\sum_{i=1}^{N'} \Omega_2^{i} + \\ \Pi_F.p_{f-f'}.\sum_{i=1}^{N'} \left( \Omega_2^{i} - \Omega_3^{i'} \right) + \\ (1-\Pi_F).p_{f-c}.\sum_{i=1}^{N''} \Omega_{f-c,t}^{i} \end{bmatrix} + P_{f}^{comp.} + P_{c}^{comp.} \end{aligned} \quad (11)$$

where, the terms inside square braces denotes power consumption at the network while transmitting the byte-stream. The second and third addend in (11) represents power consumption profile for FCNs and cloud servers respectively. The energy consumption profile of a typical fog node ($P_f^{comp}$) can be represented as a quadratic, monotonic increasing and strictly convex function of computation volume $y_i$. The function satisfies the fact that marginal power consumption of fog nodes increases proportionally with time. It also ensures that computation power consumption to be proportional to the amount of analytics activities performed. Thus,

$$P_f^{comp.} = BV_2.(p_f^{comp.}.\psi_f.\sum_{i=t-\tau}^{t} (a_i y_i^2 + b_i y_i + c_i)) \quad (12)$$

Similarly, if each data center is assumed to host homogenous computing elements (machines) of identical CPU frequency $\eta$, the power consumption component $P_c^{comp}$ of each computational element at cloud server can be approximated as a function of $\eta$, given by:

$$\mathrm{P}_c^{comp.} = (1 - BV_2).BV_{c,i}.n_{m,i}.(A_i\eta_i^{\Delta} + B_i) \quad (13)$$

where $A_i$ and $B_i$ are positive constraints. In realistic scenarios, $\Delta$ varies between 2.5 and 3. All three attributes can be obtained by curve fitting against empirical measurements when profiling the system offline [21]. The net power consumption across the architecture given by (11) is mapped to corresponding cost term $C_f^{cons}$ through energy to cost conversion parameter $\alpha_{cons}$. Thus

$$C_f^{cons.} = \alpha_{cons.}.\mathrm{P}_{net,t}^{cons.} \quad (14)$$

$$C_f^{ems.} = C_c^{ems.} = (1 - \Pi_F).\zeta.R.PUE.\beta_c \quad (15)$$

Equation (15) calculates the cost term due to emission given in (3), in terms of cost of carbon footprint $\zeta$ (USD per gram) and the average carbon emission rate $R$ found from weighted contribution of different fuel types (gram per KWh).

*C. Problem Statement*

In order to have an optimal and cost preserving FC model, it is indispensable to investigate the constraints pertaining to user-FCN association, workload allocation and VM deployment towards cost-efficient fog computing architecture. To have further insight, let us consider an illustrative connected vehicular network shown in figure 3(a) and 3(b). The toy example describes how the QoS parameters such as the upload latency, processing latency and communication latencies etc. are improved when the mode of task association/distribution, virtual machine deployment, resource allocation/association are altered. The vehicular network (VANET) shown in Fig 3(a) and 3(b) comprises of four electric vehicles having each driver using a smart charging app (single application) that recommends the optimal location of an electric vehicle charging station (EVCS). The recommendation criteria may be on economic power tariff, lowest queuing delay, shortest distance, or any/all combination of these. There are ten roadside units (RSU) each having 4 bandwidth units (BU) each. Each RSU can contribute for an uploading data rate of one per time unit and are deployed across the VANET.

The RSUs fetch different attributes, behavior, and location etc. of the commuters as input, process it to regulate the fleet dynamics. All RSUs are also interconnected through WSN or wired links having five unit communication delay between each neighboring pair. Each RSU charges 2 per time unit for hosting one VM. The RSUs also incurs an application uploading cost. In our example we assume RSUs R1, R2, R5 and R10 have uplink cost of 5, 2, 3 and 5 units respectively. For case 1, as shown in Fig. 3(a) we assume that all the vehicles are associated to R1 and data requests from all four drivers are uploaded through R1 and then processed in R1, R3, R4 and R10 respectively. RSUs R2, R3, R4 and R7 are used to host virtual machines for user applications U1, U2, U3 and U4 respectively. Our aim is to minimize the total unit cost that includes cost due to total VM deployment, queuing delay (if any), application/request uploading and inter RSU communication.

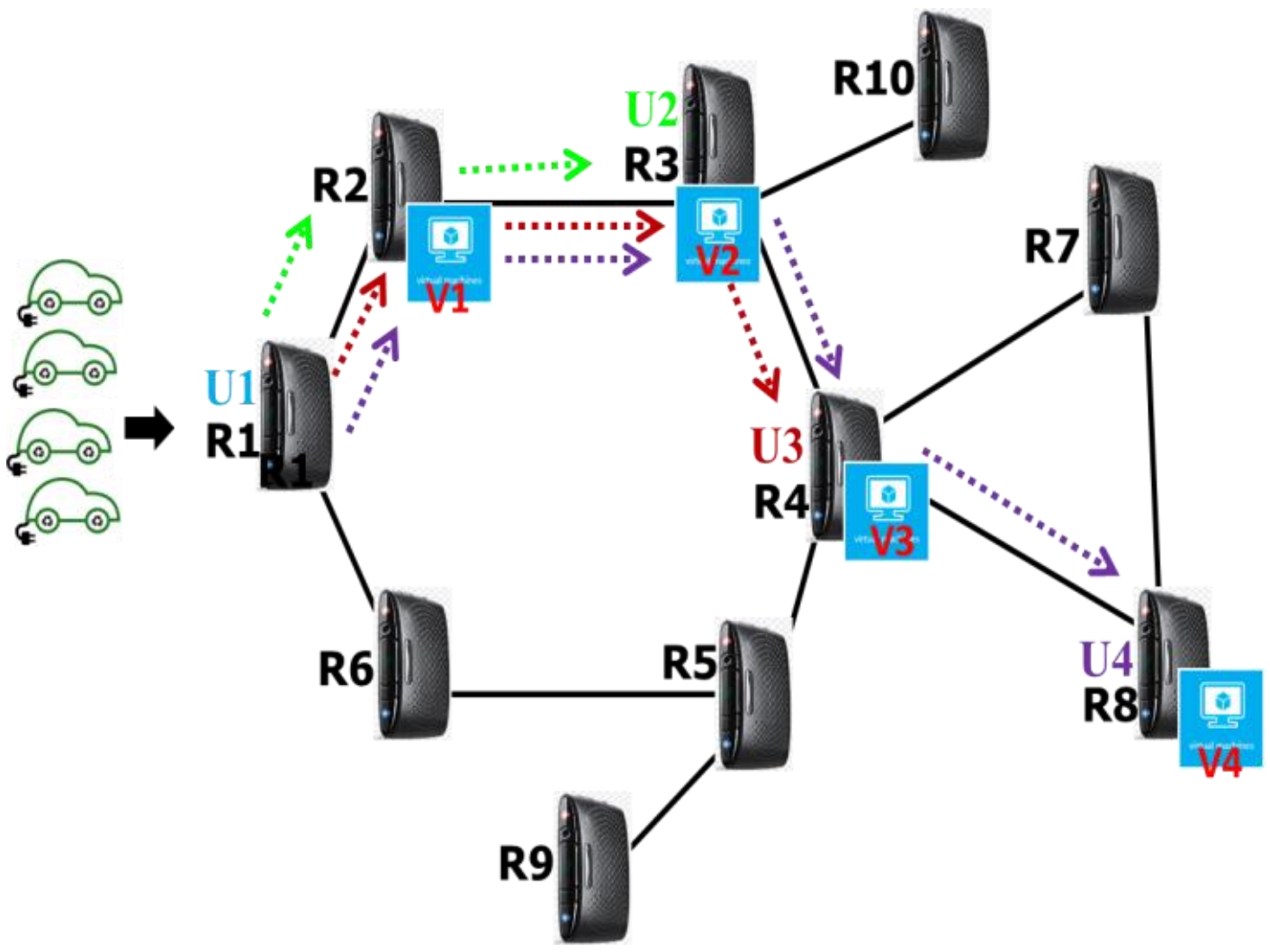

**Figure 2(a): Task association, & VM Deployment Scenario 1**

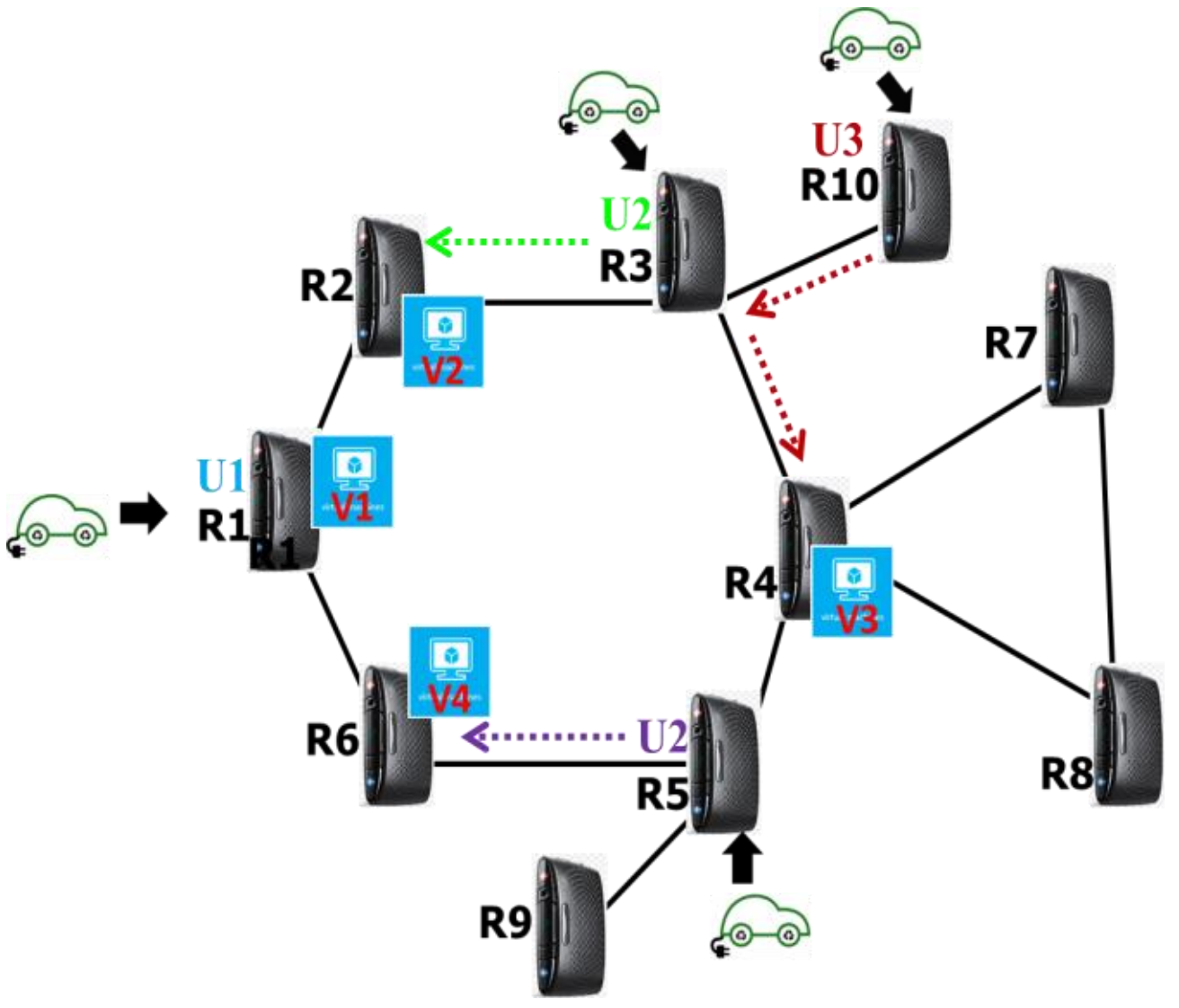

**Figure 3 (b): Task association, VM Deployment Scenario 2**

Thus the uploading cost for Fig 3(a) is 5+5+5+5=20, whereas the VM deployment cost and cost due to inter-RSU communication and total cost are (2+2+2+2==8) and (0+2x5+3x5+4x5=45) and (20+8+45=73) respectively. The uploading latency incurred by each driver in this case is 1 because each application can use only one BU of R1. If there would have less number of BUs available to each user there would be queuing delay that further raises the latencies. Now let us reconsider the network dynamics by incorporating some changes in the user-RSU association, VM deployments and resource (BU) allocation, as shown in Fig 3(b). Suppose that R1, R2, R4 and R6 are used to host the VM for applications U1, U2, U3 and U4 respectively. The respective users are now associated to R1, R3, R10 and R5. Keeping the remaining network attributes and the QoS parameters intact the cost due to request uploading, VM deployment, inter-RSU communication and total cost are (5+2+3+5=15), (2+2+2+2==8), (0+5+5x2+5=20) and (15+8+20=43). The uploading delay of all users is now decreased to 1/4 thanks to full occupancy of all four BU of corresponding RSUs. It can be inferred from this illustration that the network/infrastructure resources and the QoS can be significantly improved (here 41.09% of total cost improvement and 75% reduction in uploading delay) by changing the task association/distribution, resource allocation, computing machines deployment strategies. Thus, an optimal solution of user to Fog Computing Node (FCN) association, BU allocation, VM deployment and workload distribution not only contributes to lower total cost but also improves Quality of Experience (QoE) of applications (e.g., the response time). It is significant to investigate these factors for QoS-guaranteed minimum-cost SG applications in fog computing environments. The optimization framework proposed in this work includes an extensive range of constraints to investigate base station association, task distribution, and virtual machine placement to ensure an optimal QoS for a fog archetype in SG.

In this section, we present an MINLP formulation on the minimum cost problem with the joint consideration of data consumer association, workload allocation, VM deployment and Networking constraints for communication infrastructure.

*1) User-FCN association constraints:*

Since the SG have both mix of static as well as dynamic data consumers, at each time epochs the FCN maintains a dynamic list $L$ to track the reachable data consumers, user $i$ can only be associated to FCN $j$ only if $j$ is reachable from $i$ . In other words, FCN $j$ is reachable from query generation source node $k$ if the former is in the preference list $L$.

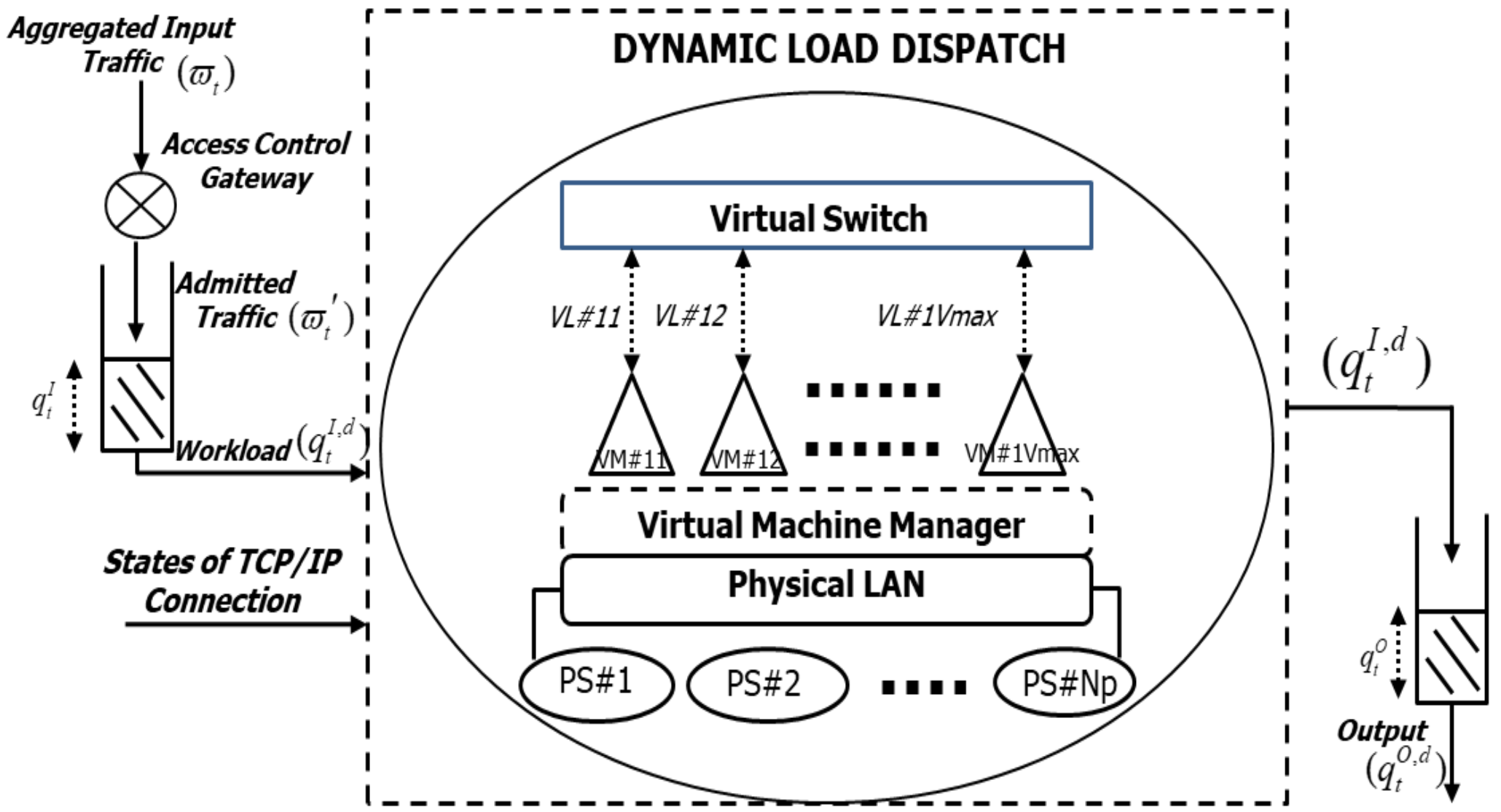

**Fig. 4: General architecture of a virtualized Fog Computing Node (FCN). It operates at the Middleware layer. (PS = Physical Server, VL= Virtual Link, VM=Virtual Machine).**

Therefore,

$$\sum_{a \in A} BV_3 \leq BV_L \tag{16}$$

Where, $BV_L$ is a binary variable indicating if user node i is in the list $L$. Also, an FCN $j$ can associate user i only when it has atleast one available bandwidth units (BU). If $S$ denotes the set of BUs corresponding to FCN $j$, then

$$\frac{\sum BV_3}{|S|} \leq BV_2 \leq \sum BV_3 \tag{17}$$

Since there is no restriction on the number of BUs allocated to a consumer from a fog node; but one BU can be only allocated to at most one consumer, i.e. there is a many one mapping from each set of BU to those of FCN, i.e.

$$\sum_{a \in A} \boldsymbol{BV_3} \leq \mathbf{1} \tag{18}$$

In order to granulate the QoS of applications from whole range of users, one premise is that every consumer must be associated with one and only one FCN, i.e.

$$\sum_{f \in F, a \in A} BV_2 = 1 \tag{19}$$

*2) Workload allocation constraints:*

Since the data streams may undergo sequential fog processing, constraint (20) indicates that fog node $f$ can distribute application data to processing at fog node $f'$ only when it is associated with application consumers in the list $L'$ of $f'$. Thus,

$$BV_4 \leq \sum_{a \in A} BV_2 \tag{20}$$

If $\lambda^{a}_{f-f'}$ denotes inter-fog (*f-f ′*) data arrival rate, then

$$\frac{\lambda^{a}_{f-f'}}{\mathbb{R}} \leq BV_4 \leq \lambda^{a}_{f-f'}.\mathbb{R} \qquad (21)$$

Corresponding to each application request, if the task need to be accomplished by more than one fog nodes, constraint (22) ensures that all uploaded data are completely processed and total data received from data consumers through fog node *f* shall be equal to data finally processed at all fog node *f ′*.

$$\sum_{a\in A} r^{a}_{f-f'}.BV_2 = \sum_{f'\in F} \lambda^{a}_{f-f'} \qquad (22)$$

*3) VM deployment constraints:*

In principle, the FCN may serve the connected IoT devices according to all three general models, namely, the Infrastructure as a Service (IaaS), Platform as a Service (PaaS), and Software as a Service (SaaS) [25]. However, due to the stringent hardware and software resources limitations of the of the IoT devices in SG environments, only the SaaS model seems to be the most feasible one. According to Cisco [26], a state-of-the-art FCN is an IoE-compliant SaaS-oriented software, that aims at mapping raw sensor data into actionable information. Roughly speaking, an FCN defined for our purpose is composed of five components as shown in Fig. 4.

(a) **Admission Control Router (ACR):** For defining the admission control policy, that maintains loss-free input and output queue.
(b) **Input buffer:** A time-slotted $G/G/1/N_I$ FIFO queuing system for modelling the input queue.
(c) **Output buffer:** A time-slotted $G/G/1/N_O$ FIFO queuing system for modelling the output queue.
(d) **VLAN:** It consists of a reconfigurable computing platform and the related switched Virtual LAN
(e) **Adaptive scheduler:** It reconfigures and consolidates the available computing-communication resources and also performs the control of the input/output traffic flows.

Let each fog node (FCN) is equipped with $N_P$ physical servers**.** Each server can host at the most $\mathbf{V_{max}}$ number of VMs. Thus the maximum number of VMs hosted by an FCN *i* is given by:

$$V^{i}_{M} = \sum_{i=1}^{N_P} V^{i}_{\max} \quad \forall\, i = 1, 2, \ldots N_P \qquad (23)$$

In a virtualized FCN, each VM processes the currently assigned workload by self-managing own local virtualized storage and computing resources. When a request for a new job is submitted to the FCN, the corresponding resource scheduler adaptively performs both admission control and allocation of the available virtual resources. At the end of time slot *t*, new input requests arrive at the input of the ACR (Fig.. 4) following Poisson distribution. For $t \geq o$, the random process $\varpi_t$ defining the arrival pattern is presumed to be independent from the current backlogs of the input/output queues. At the end of slot t, it is assumed that any new arriving packet that is not admitted by the ACR is declined. Thus, out of $\varpi_t$ if $\varpi_t'$ is the number of BUs per slot, that are admitted into the input queue, then

$$0 \leq \varpi_t' \leq \varpi_t \qquad (24)$$

The respective time evolutions of the backlogs in input and output queues is given by

$$q_{t+1}^{I} = \left(q_t^{I} - q_t^{I,d}\right) + \varpi_t' \qquad (25)$$

$$q_{t+1}^{O} = \left(q_t^{O} - q_t^{O,d}\right) + q_t^{I,d} \qquad (26)$$

For the data stream generated by application $a$ to be processed in FCN $f'$, i.e. $BV_4 = 1$ and corresponding VM must be deployed in *k* and vice versa. The following constraints are defined respectively for that purpose.

$$\frac{r_i^s}{\mathbb{R}} \leq BV_5 \leq r_i^s \mathbb{R} \qquad (27)$$

$$BV_4 \leq BV_5 \qquad (28)$$

if $Q_a^s$ , $Q_f^s$ and $Q_f^p$ respectively denote the resource requirements of any application $a$ and the storage capacity of FCN *f*, and its processing capability, then equation (25) and (26) say that requirements of any application at any time is limited by storage and computational capacity of deployed VM, i.e.

$$\sum_{a \in A} BV_5 . Q_a^s \leq Q_f^s \qquad (29)$$

$$\sum_{a \in A} r_i^s . \varepsilon^a \leq Q_f^p \qquad (30)$$

where, $\varepsilon^a$ is a scaling factor to indicate the relationship between processing speed and allocated computation resource.

*4) Network constraints for FC:*

In addition to constraints (5)-(8), at any instant, constraint (27) maintains the stable queuing equilibrium by keeping the service request rate less than the service rate. Constraint (28) ensures that the expected delay for any application or consumer at any instant should not exceed the specified delay limit imposed by QoS requirements.

$$\sum_{f' \in F} \lambda_{f-f'}^{a} < r_i^s \qquad (31)$$

$$L_{fc}^{a} = L_{f,i}^{upd} + L_{f,i}^{comp.} + L_{c,f2f}^{comm.} + L_{f-c}^{dispatch} + L_{c}^{comp} < L^{a} \qquad (32)$$

## *D. Optimization Model*

In this section, we present an MINLP formulation on the minimum cost problem with the joint consideration of data consumer association, workload allocation, VM deployment and Networking constraints for communication infrastructure. The objective of the proposed framework is to minimize the overall cost incurred due to power consumption, latency and carbon footprint. By summarizing the constraints discussed above, the cost optimization problem can be formulated as a mixed-integer nonlinear programming (MINLP) problem, i.e.

$$Maximize \left\{\min C_F^T - \min C_C^T\right\} \qquad (33) \qquad s.t.\ \textit{Constraints}\ \ (16)\text{ to }(32)$$

# V. Simulation and Algorithmic Set-up

The essential nodes in the system include set of cloud servers D, fog nodes F and data users N. The architecture is virtually deployed on the essential nodes in an arbitrary SG network. For simulating the pilot SG topology, the 100 most populated places around the world are considered (i.e. $|F| = 100$), the corresponding population for representing the number of consumer/data generation nodes and the corresponding geographical coordinates are used to determine the relative Euclidian distance [27]. The consumer endpoints within a particular city are logically grouped to form a cluster and are associated to an FCN. The generated data traffic will be proportional to the population of Internet users of the corresponding city. The number of servers across the globe is considered to be 8, the pairwise Euclidian distance is stored in 2D variable $D_E[\,][\,]$, when their geo-location is determined through clustering of city population.

Each instruction is of size 64bits. The user to fog links allows transmission of packets of 34 to 64K bytes following Poisson arrival pattern having 8 byte instruction size and having mean packet arrival rate being 1 packet per node per second. The capacity of the links between data generators and FCNs is considered to be 1 Gbps whereas the inter-fog communication link capacity is taken as 10 Gbps. However, WAN communication between the FCNs and the cloud servers are assumed to take place through bandwidth unconstrained channels. The total number of data consumers in the system is treated as a variable, within the range $[10^4, 10^5]$, to assess the system performance against varying network conditions. The cloud servers transmit their data through access points distributed across each SG network. Each homogenous cloud server is assumed to accommodate the varied number of IoT devices within the discrete set {16K, 32K, 64K, 128K}, based on the network traffic to be processed. Energy consumption rate of each FCN is taken to be 3.7W while for cloud servers it is taken to be proportional to the number of IoT devices associated to each of them and taken from the range {9.7, 19.4, 38.7, 77.4} MW. The cost corresponding to consumed power is uniformly distributed between USD 30/MW h and USD $70/MW h [28]. For cost analysis, the cost of 1Gbps and 10Gbps gateway router port is kept USD 50 each per year while cost of server is USD 4000 per year [28]. These routers are assumed to consume electricity at 20W and 40W respectively. Upload tariff is USD 12 per byte while storage cost is kept in the range of USD 0.45-0.55 per hour. The penalty corresponding to $CO_2$ emission is kept USD 1000 per tons of $CO_2$ emitted [29]. In order to obtain an optimal value of probability $(\pi_C)$ (best estimates that maximizes fog utility in (16)), Monte Carlo (MC) simulations are performed, having number of trials set to (1000, $T_{MC}$), where $T_{MC}$ is the number of required MC trials that ensure a 95% confidence interval of an error less than 1% [30]. Further to improve the efficiency of scenario generated through MC, Latin Hypercube Sampling (LHS) [31] is used. LHS is a low discrepancy technique that generates evenly distributed random samples with small variance.

The formulated optimization model is a multistage, discrete, ***non-convex***, constrained mixed-integer non-linear programming problem (MINLP). Usually classical mathematical programming techniques fail to provide tractable solutions to such problems. Evolutionary optimization algorithms specifically meta-heuristic methods such as differential evolution (DE) [31] provide promising approach to solve an MINLP. DE is a population-based evolutionary optimization method which had proven to be very simple yet powerful to solve minimization problems with nonlinear and multi-modal objective functions. It differs from conventional evolutionary algorithms in that instead of having a predefined probability distribution function (*pdf*) for mutation process, it utilizes the differences of randomly sampled pairs of objective vectors for its mutation process [32]. Such variations will ensemble the topology of the objective function towards optimization procedure thus providing more efficacious global optimization capability. In order to reduce the fitness of similar offspring's, we employed, a modified version of differential evolution with a fitness sharing function of niche radius ($\rho$). For each individual population *i* and given a threshold value of niche radius $\rho$, the DE calculates the shared fitness $S_f'$ according to (34). Calculating the shared fitness $S_f'$ before selection operation supplements significant computations for evaluating values before executing selection operation. The underlying principle of employing fitness sharing is to cluster the population into smaller groups defined by a similarity measure. In this work the similarity is defined by a distance function $\mathbf{d_{ij}}$ that satisfies equation (30).

$$f(\mathbf{d}_{\mathbf{ij}}) = \begin{cases} 0 \le f(\mathbf{d}_{\mathbf{ij}}) \le 1 \\ f(0) = 1 \\ \lim_{\mathbf{d}_{\mathbf{ij}} \to \infty} f(\mathbf{d}_{\mathbf{ij}}) = 0 \end{cases} \tag{34}$$

The individuals which lie in same group will share the corresponding fitness value and in the selection operation, clusters having larger fitness sharing value will be selected for producing the next generation offsprings. The fitness sharing function is given by:

$$f(\varphi, \mathbf{d_{ij}}) = \begin{cases} 1 - \left( \dfrac{\mathbf{d_{ij}}}{\rho} \right)^{\varphi} & 0 \le \mathbf{d_{ij}} < \rho \\ 0 & otherwise \end{cases} \tag{35}$$

The niche count $\rho_c$ for every individual $i$ can be calculated as $\rho_c = \sum_{j=1}^{F} f(\varphi, \mathbf{d_{ij}})$ (36)

**Algorithm 1**

1. ***Initialize*** main population set $\mathbf{P}$ with $|\mathbf{P}| = F$
2. ***Evaluate*** the population $\mathbf{P}$
3. ***Copy*** elite solutions to elite set $\mathbf{E}$ with $|\mathbf{E}| = F/2$
4. **While** stopping criteria not met **do**

- {*Generate* hybrid population $\mathbf{H}$ with $|\mathbf{H}| = F$
- For each individual $i$ in the population,

  ***Find*** $S'_f$ using (34)

- $\forall i \in \mathbf{H}$ **do {**
  - Descending sort on $\mathbf{H}$ based on Shared fitness $S'_f$
  - Remove the worst half of $\mathbf{H}$

  **} While** (mixed population is full)

- **do** $\mathbf{H}' \leftarrow \mathbf{H} + \mathbf{E}$ where $\mathbf{H}'$ is the combined population set.
- **do** descending sort on $\mathbf{H}'$
- Copy first $F/2$ elite solution from $\mathbf{H}'$ to form new elite class $\mathbf{E}'$
- **do** $\mathbf{P} \leftarrow \mathbf{H}$

Since during selection process, individuals with large fitness survive and used for mutation or recombination purpose, the value of shared fitness can be calculated by:

$$S_f = \frac{f_j}{\sum_{j=1}^{F} f^{j}(\varphi, \mathbf{d_{ij}})} \tag{37}$$

However, for our purpose, since the evolution strategy is focused to obtain minimal optimal value, the shared fitness $S_f'$ is obtained from the equation

$$S'_f = f_j \cdot \sum_{j=1}^{F} f^{j}(\varphi, \mathbf{d}_{ij}) \quad (38)$$

Where, $f_j$ controls the crossover constant commonly determined on a case to case basis having $\varphi$ as the control parameter and often set as $\varphi = 1$. .

In order to guarantee the fact that the best offspring always appear for next generation, elitism is employed. Algorithm 1 shows the progress of modified DE employed. When the BUs are exclusively allocated to an application user, the update latency cost is determined by the uplink rate $\delta$ (see equation. 5)), regardless of the data volume. Similarly, the inter-FCN communication cost and WAN dispatch latency is determined respectively by the actual network traffic $\ell_c . r_{i,j}$ (where $r_{i,j}$ is the request rate for application $a$ from FCN $i$ to FCN $j$) and WAN communication bandwidth $\omega_c$. The overall goal is to maximize utility function (33) having optimal settings of $\pi_C$ $\pi_F$, $BV_1 - BV_5$ $BV_L$ and $r_i^s$. The decision variables for (33) are the workload $r_{i,f}^a$ assigned from user $i$ to FCN $f$ , workload $\boldsymbol{y_j}$ dispatched from to FCNs to data center $j$, the traffic rate $\Gamma_{f-c}$ dispatched from FCN $f$ to data center $c$, CPU frequency $\eta$ of homogenous servers installed at these data centers and the number of turned-on machines $n_{m,i}$ on server $i$. After applying MDE algorithm we obtain optimal workloads $\left(r_{i,f}^a\right)^+$ and $y_j^*$ . Correspondingly power consumption, latencies and total cost of architecture for both fog and cloud scale processing can be calculated.

## VI. PERFORMANCE EVALUATION

In this section we presented a comprehensive comparison of cloud and fog scale execution in terms of performance metrics namely response times (service delay), electricity consumption and cost of architecture. We depict the overall latency profile of fog assisted cloud architecture with a generic cloud execution scenario. The upload latency, inter-fog communication delay and delay due to fog to cloud dispatch is abstracted in transmission latency term while the delay caused due to computations and analytics at VM fog nodes and cloud servers is covered under processing latency term. The overall response time (service delay) is given by algebraic sum of transmission latency and processing latency. We define Fog Network Efficiency (FNE) $\Re$ (equation 39) as the ratio of data packets dispatched to cloud core to the number of packets entering into the fog network through consumer to fog gateways. Higher value of $\Re$ indicates more fraction of applications demand realtime responses.

$$\Re = \frac{\mathbb{N}_C}{\mathbb{N}_F} = \frac{\mathbb{N}_F - \mathbb{N}_f}{\mathbb{N}_F} \quad (39)$$

For instance, $\Re = 0.8$ means only 20% of consumer requests demand fog plus cloud scale computations and is better than $\Re = 0.01$ because in that case almost (99%) of data traffic needs cloud scale processing also. Such scenario extrapolates to pure cloud paradigm (even worse if response time is the only QoS of the system) for very low magnitudes of $\Re$. For $\Re = 1$, the offloading model degenerates to pure edge computing paradigm.

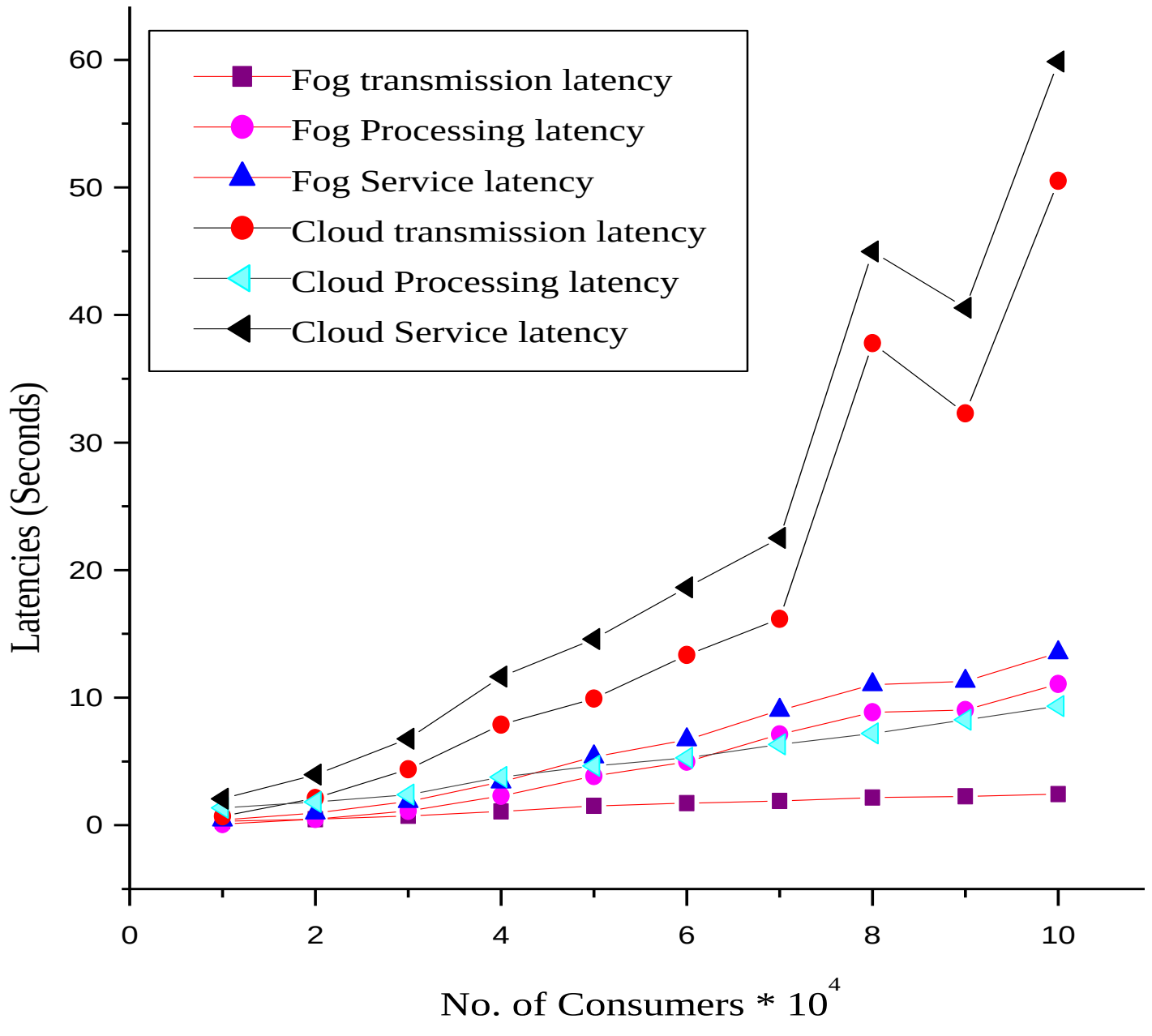

**Figure 5: Comparison of Latencies in Fog and Generic Cloud Computing Models**

The fog-cloud delay statistics is shown in Fig. 5 for $\Re = 0.25$ (three fourth of requests are served within fog alone). The mean transmission latency mean processing latency and service latency are plotted, separately for both cloud and fog platforms, against variable number of consumer nodes. It can be observed that for both the fog as well as cloud platforms the latency is proportional to the density of data generators (traffic). Meanwhile the performance of fog assisted cloud execution outperforms the cloud counterparts for every magnitude of data traffic.

**Table 1: Change in Execution Strategy with varying FNE**

| Rate | $\Re$ | Execution Model |
|---|---|---|
| **1** | 1 | Edge |
| **2** | 0.8 | Fog assisted cloud |
| **3** | 0.5 | Fog assisted cloud |
| **4** | 0.01 | Fog assisted cloud |
| **5** | 0 | Pure cloud |

We consider the changes of the magnitude of Fog Network Efficiency $\Re$ defined in (39) in the range [0, 1] as shown in Table 1, and plot the transmission latency and processing latency, and observe the change in the corresponding service latencies in Fig. 6. It can be observed that as the value of $\Re$ is scaled from .05 (5%) to higher magnitudes, the response time is significantly improved i.e. the overall network latencies (average transmission latency 6(a), average processing latency 6(b) and average service latency 6(c) ) is reduced.

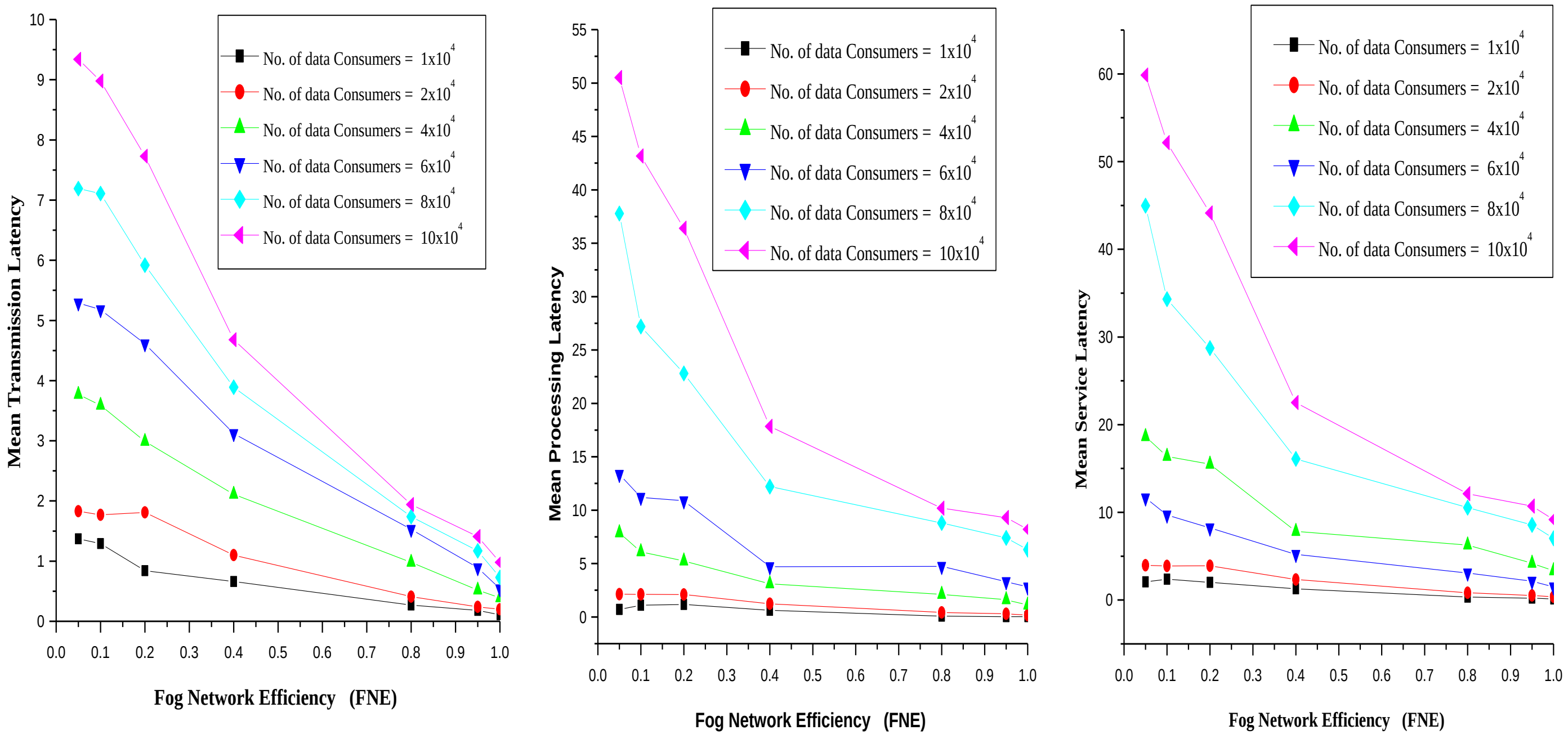

**Figure 6 (a)-(c): Variation of latency parameters against Fog Network Efficiency (FNE)**

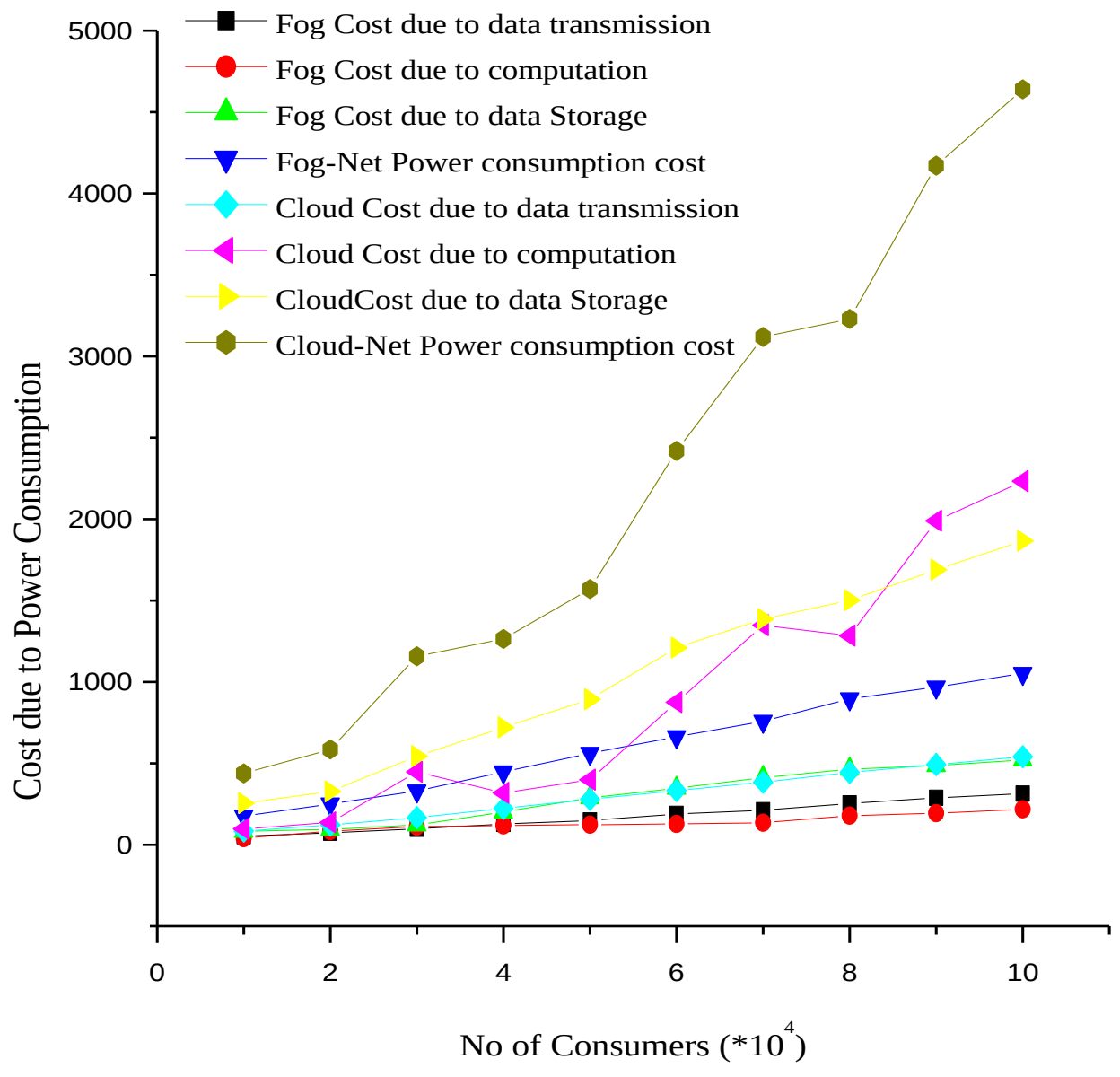

**Figure 7: Comparison of various cost parameters involved in Fog and Generic cloud computing**

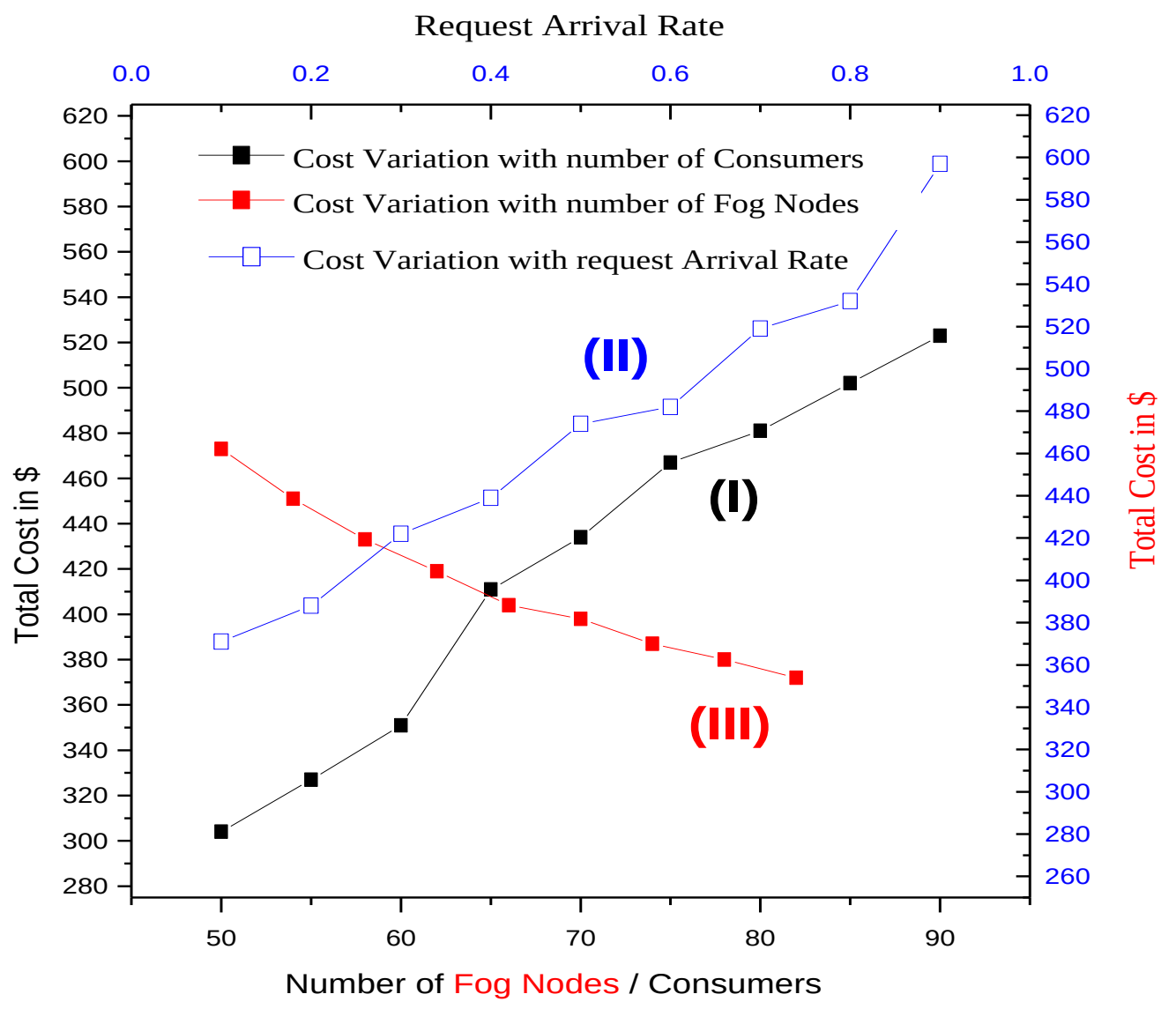

**Figure 8: Cost variation with varying attributes**

Fig. 6(a)-(c) shows the most significant observation of the proposed model that, with the increase in the magnitude of $\Re$ i.e. as more number of application requests demand real-time and latency-sensitive services (workload percentage on the FCNs increases), the mean transmission latency and the mean processing latency are significantly reduced. For an infrastructure with $\approx$ 50% applications requesting real-time services (i.e. $\Re \approx 0.5$), the overall service latency for fog computing is noted to get reduced to almost half of that of pure cloud paradigm. Also, with the increase in the number of workload generators (consumer

nodes), the latencies increase. For lower number of consumer nodes and at low values of $\mathfrak{R}$ (<0.13 in 6(a)-(c), red and black colored curves), the latencies are almost same for both cloud and fog platforms. This indicates that adopting fog computing is viable only when the data traffic is huge (Big-Data Analytics) and is not economic for small scale computations. Also, in the context of IoT aided environment such as SG, if the percentage of applications demanding for real-time services is low, then fog computing may come with an overhead compared to the traditional cloud computing.

In Fig. 7 the electricity consumption pattern due to transmission/dispatch of each bytes-streams and computation (at both fog and cloud servers) is analyzed. It can be observed that with the rise in the population of service consumers the overall power consumption show near piece-wise linear growth and is significantly lower than the conventional cloud framework. The fog assisted cloud framework betters the aggregated electricity consumption over the cloud computing paradigm by more than 40%.

Fig 8 shows the variation of cost when the network parameters are varied i.e. when the number of application users, the data arrival rate and number of fog nodes etc. For Fig. 8, the optimization model was run on a pilot network of only eighty users and fifty FCNs having five BUs allocated each fog nodes. Curve (I) shows the variation in the cost of architecture when the number of consumers are varied from 50 to 95. It is observed that the cost profile also shows a nearly piece-wise linear growth corresponding to rise in the population of fog customers. This supports the intuition that more customers will create more service requests, thus generating more data traffic and hence more VMs need to be deployed. The cost variation against change in arrival rate is shown in curve II where the architecture also shows similar cost profile because in order to guarantee optimal QoS more BUs as well as processing resource are needed. Correspondingly the cost due to communication latencies ($C_f^{comm}$) also increases, thus augmenting the overall cost. However, if more and more fog nodes are deployed more efficiently the task will be accomplished and correspondingly better options for VM deployment. Similarly curve (III) illustrates the total cost as a decreasing function of the number of Fog Nodes. The cost of task distribution and virtual machine deployment algorithms decrease significantly when the network is populated with more number of FCNs, hence the total cost decreases.

## VII. RELATED WORKS

### *A. Works related to Fog Computing in Smart Grid*

In future smart cities, smart power grids will be critical in ensuring reliability, availability, and efficiency in city-wide power management [33]. A successful SG architecture will be able to help improve transmission efficiency of electricity, react and restore timely after power disturbances, reduce operation and management costs, better integrate renewable energy systems, effectively save energy for future usage. This vision is bold but critical to enabling smart living [34]. The new challenges of big data analytics (BDA) posed by SG welcome scientists from both academia and R&Ds to investigate and develop novel and high-performance computing architectures [11]. Due to its multi-faceted opportunities and advantages over pure cloud and edge models, fog computing had received enormous attention from the SG community. In fact in the seminal paper where the concept of FC was first proposed, the authors presented the SG architecture to be a rich use-case of fog computing [23]. They further enlightened the scope of fog based SG deployment taking a micro-grid (wind farm) example in [35]. An FC based SG prototype is presented in [36], where the authors examined how the FC environment can act as a bridge between the SG and back end cloud for offering store and compute services for smart meter from AMI. Through smart home example, they show how an FC based SG deployment will assist the customers to monitor, analyze and fine-tune their daily/weekly/monthly electric consumption in a secure manner. However, sound mathematical foundation was not drawn to support the model. Besides, a range of literatures [3], [5], [33], [37], [38] exemplify the sprouting opportunities, challenges and research directions in fog based SG and emphasize the significance of software platforms as well as SOAs for big data analytics in SG.

The SG transportation networks must adapt to dynamic usage, traffic conditions, and user behavior with a minimal carbon footprint. A clean and renewable power grid must actuate localized energy and power control. Also, pervasive security enforcement is needed to detect and prevent potential threats [34]. Jalali et.al. in [39] proposed fog computing as a technology enabler for green IoT. They compared the power consumption of multiple IoT applications when running on both Fog and Cloud platforms. However they considered only energy metric for comparison purpose, not the full range of metrics that defines a typical fog environment. In [38], the authors took the instance of Los Angeles Smart Grid, the largest public utility in the US that will serve over 4 Million electricity consumers, to highlight the scope of fog based deployments. They hypothesized a test bed where the IoT integrated smart meters from AMI running over P2P or M2M communication channels will observe energy demand at households and industries and report them periodically back to the utility every few minutes. For demand response operations, the gateways can act as fog nodes take local decisions to determine curtailment strategies and control, say a smart appliance or an electric car, or centrally change set-points of HVAC systems across campus buildings [6]. The suitability of fog computing is also explored for the case of distributed state estimation models, but they nowhere present any experimental framework to validate the proposals [37], [38].

### *B. Works related to Workload Allocation in Fog Computing*

Inspection of the recently published works support the conclusion that resource provisioning and workload allocation in fog assisted cloud architectures is still in infant phase and still lacks concrete solutions to reveal its viability. In [50], scheduling based workload allocation policy is presented to balance computation load on FCNs and client/user devices. The authors in [40], propose a workload allocation framework for optimizing the delay and power consumption in Fog-Cloud interaction. In [41], the authors focus on task scheduling algorithms for the minimization of the energy in reconfigurable data centers that serve static clients. They proposed a greedy strategy based scheduling algorithm for mapping tasks to VMs and then to suitable servers. Though the achieved energy performance is appreciable but their usecase does not consider mobile clients, which are pervasive in SG. Similarly, the authors in [42] present a programming model including a simple resource provisioning strategy, which relies on workload thresholds, i.e., if the utilization of a particular fog cell exceeds a predefined value, another fog cell is leased. Apart from fog-specific resource provisioning solutions, resource allocation and service scheduling are major research challenges in the general field of cloud computing [43], [44]. Though such tactics offer motivating insights, there are key differences between fog services and cloud services. Thus it prevents a direct adaptation for the use in the work at hand. First, the size and type of fog resources are very different from its cloud computing counterparts. While cloud resources are usually handled on the level of physical machines, virtual machines (VMs), or containers, fog resources are usually not as powerful and extensive. While cloud resources are usually placed in centralized data centers, the FCNs may be distributed in a rather wider area having heterogeneous network topology, making it more important to take into account data transfer times and cost in FC. This is especially important since one particular reason to use FC in IoT scenarios is the higher delay-sensitivity of fog-based computation. Hence, resource provisioning approaches for the fog need to make sure that this benefit is not foiled by extensive data transfer times and cost. Resource and workload allocation fog like networks may be performed through optimization algorithms. One such approach could be seen in [45], where the authors define an iterative optimization algorithm based on weighted spherical means. We found analogous efforts in healthcare applications [46], where the objective is to solve the problem of the unstable and long-delay links between the cloud data centers and medical device(s). A case study on fog based Electrocardiogram (ECG) feature extraction is performed in [47], to diagnose cardiac diseases. Interestingly, more than 90% bandwidth efficiency was achieved for the same but they did not consider other QoS metrics such as energy consumption or carbon footprint etc. A prototypical Smart e-Health Gateway or fog node called UT-GATE was devised in [48] for IoT based Early Warning Score (EWS) application. Extensive simulations were presented to demonstrate the enhanced overall system intelligence, energy efficiency, mobility, performance, interoperability, security, and reliability. But to the best of our knowledge, we found no such contribution focusing on workload allocation, resource-provisioning and feasibility analysis of fog based SG platforms. Inspired by such concept(s), our work further investigates a fog based SG architecture coupled with a cost-efficient workload allocation framework and provides rigorous theoretical results to guide the practical SG deployment.

## VIII. CONCLUSION AND FUTURE WORK

Fog Computing when complemented with optimal workload allocation strategies, is able to support context-aware virtually real-time applications, providing improved computing performance, and geo-distributed intelligence in Smart Grid ecosystems. In this work, we presented a fog-based data intensive analytics scheme with cost-efficient resource provisioning optimization approach that can be used for SG applications. In order to achieve QoS guaranteed FC execution strategies, we jointly examine user-FCN association, workload allocation, VM deployment and communication network constraints towards minimizing the cost of architecture. For comparative performance assessment of cloud versus fog computing, we formulated an MINLP optimization problem, which is further solved using MDE algorithm. Exhaustive simulation results are presented to depict the enhanced performance of FC in terms of the provisioned QoS attributes viz. service latency (response time), power consumption and cost of architecture.

Such observations related to FC reveal its latent potential as a store & compute model for emerging IoE environments such as SG. For further conclusive insights, real-world trials need to be conducted on SG architectures. The efficacy of typical FC paradigm may be improved by embedding selective sensing intelligence in the edge nodes as well dedicated FCNs. Intelligent mobility management techniques for data generators and data consumers will potentially improve the performance of FC architectures.

The current FC deployment due to its nature faces new security and privacy challenges. The existing security and privacy measurements for cloud computing cannot be directly applied to the fog computing in SG like due to its features such as mobility, heterogeneity, large-scale geo-distribution. State-of-the-art authentication, privacy and security protocols, standards and solutions are key to successful FC rollout. Furthermore, orchestrating FC supported candidate applications can simplify maintenance and enhance data security and system reliability. Thus it is crucial to have efficient orchestration mechanisms to deal with the dynamic variations and transient operational behavior of those services. Rigorous study & analysis of composability and concrete methodologies need to be performed to develop prototypical orchestration frameworks for FC in SG architectures requires. Within lifecycle management of fog architectural entities, optimal selection and placement of nodes in the deployment stage, dynamic QoS monitoring and providing guarantees at runtime through incremental processing and re-planning, big data-driven analytics ($BD^2A$) and optimization approaches etc, are nascent research thrusts that will leverage utilities to improve orchestration quality and accelerate optimization for problem solving. Since FC networks are faced by redundancy issues, unifying redundancy management frameworks ensuring trade-off between computation latency, power consumption and communication load, need to be developed towards robust and reliable operation of FC architectures.